\documentstyle[aps]{revtex}

\begin{document}
\author{A. Zh. Muradyan$^{1,2}$, H. L. Haroutyunyan$^2$}
\address{$^1$Department of Physics, Yerevan State University, 1 Alex Manukian, Yerevan%
\\
Armenia\\
$^2$Engineering Center of Armenian National Academy of Sciences, Ashtarak-2,%
\\
378410 Armenia;\\
E-mail:muradyan@ec.sci.am; yndanfiz@ysu.am.}
\title{Large Scale (Multiphoton) Evolution of Atomic Energy Level's Mean Momentum
for Single-Photon Absorption/Emission Process}
\maketitle

\begin{abstract}
It is shown that a two-level atom, being initially in general superposition
state of ground and excited energy levels with mutually different momentum
distributions there, gets a large scale evolution in the energy levels
momentum distribution. As a consequence the mean momentum of each individual
energy level also gets large scale changes, more than the own momentum $%
\hbar k$ of the photon\footnote{%
Because of not completeness of the literature under the hand, we are not
sure that this simple phenomenon isn't known in quantum theory of
atom-photon interactions or, particularly, in the interferometry of atomic
matter waves. One of main purposes of this article is our request to the
leading specialists to respond and if the phenomenon is known, send the
corresponding references.}. Thoroughly is discussed the special case, when
the atom's preliminary superposition state is created as a result of
interaction of the atom with the resonant standing wave. Also it is pointed
that in such conditions the mentioned phenomenon can be presented as a
transformation of the resonant Kapitza-Dirac splitting of atomic states into
the Stern-Gerlach type splitting.
\end{abstract}

\section{Introduction}

Interaction of a two - level system with a plane traveling wave leads to
one-photon transitions between energy levels and consequent changes of total
atomic momentum in limits of one photon momentum $\hbar k$. What can be said
about distributions and mean values of momentum for each individual atomic
level? The answer is well-known and is trivial, if the atom before the
interaction is on one of the energy levels: distribution on the other level
sets shift $\hbar k$ and mean values of momentum distinguished also by $%
\hbar k$; $\overline{p_e}=\overline{p_g}+\hbar k$, where $\overline{p_g}$
and $\overline{p_e}$ are mean values of momentum on ground and excited
levels consequently ($1D$ case).What would we have in general case, that is,
when the atom before the interaction with the travelling wave is in the
superposition state of ground and excited levels with mutually different
momentum distributions there? In the following just this question will be
elucidated in details: in general form in Sec. 2, and the special case of
preparing standing wave, latter.

In Sec. 2 it will be shown that in one-photon absorption/emission process,
in general, large scale redistributions of the energy level's momenta takes
place. This entails the time-changes of the $\overline{p_g},$ $\overline{p_e}
$ and these changes can greatly exceed the value of one photon momentum. In
general, the relations $\overline{p_e}=\overline{p_g}+\hbar k$ are not true,
too.

In Sec. 3 we will discuss from the practical point of view a very important
case, when the preliminary superposition state of the atom is realized by
the coherent diffraction of the atom in the field of resonant standing wave,
which is being often referred to us the resonant Kapitza-Dirac effect. Will
be pointed out, that the redistribution of momenta in the travelling can be
received as a transition from the resonant Kapitza-Dirac splitting to the
Stern-Gerlach type splitting. In Sec. 4 will be discussed the temporal
behavior of the energy level's mean momenta. The results are summarized in
Sec. 5, where the possibility of experimental observation of this phenomenon
will be sketched too.

\section{Distributions and mean values of momentum on the ground and excited
energy levels in the Absorption/Emission of one photon}

Let's discuss the resonant interaction of a two-level atom with the
radiation field \cite{Landau}. For the sake of simplicity, suppose the field
has plane wavefront, linear polarization (these assumptions will be
conserved for the standing wave, discussed in Sec.3), turns on instantly and
after on it's amplitude remains constant. Let the wave functions of free
two-level atom's ground ($g$) and excited ($e$) levels be $\varphi _g(%
\overrightarrow{\rho },t)$ and $\varphi _e(\overrightarrow{\rho },t)$
respectively, where$\overrightarrow{\rho }$ is atomic internal coordinate
(radius-vector of the optical electron relative to the atomic
center-of-mass). The wave function of interacting atom will be \cite{Landau} 
\begin{equation}
\Psi =A\text{ }\varphi _g(\overrightarrow{\rho },t)+B\text{ }\varphi _e(%
\overrightarrow{\rho },t)  \label{1}
\end{equation}
where $A$ and $B$ are the probability amplitudes of the atom to be on the
ground and excited levels correspondingly.

When taking into account translational motion of the atomic center of mass,
it is necessary to separate respective parts (wave functions) in $A$ and $B$
coefficients. If, for example, the atom on thw energy level has the
well-defined value of momentum $p$, the respective wave function is given by
function 
\begin{equation}
\chi (p)=\frac 1{\sqrt{2\pi \hbar }}\exp (\frac i\hbar p\text{ }z),
\label{2}
\end{equation}
that is, by the exponential function with imaginarily degree. In general,
when on energy levels an atom hasn't definite values of momentum, $A$ and $B$
coefficients will be expressed by the series of $\chi (p)$-states: 
\begin{equation}
A(t,z)=\int a(p,t)\chi (p)dp,\text{ \quad }B(t,z)=\int b(p,t)\chi (p)dp,
\label{3}
\end{equation}
with probability amplitudes $a(p,t)$ and $b(p,t)$ of atom to have momentum $%
p $ (at the $t$ moment of time), simultaneously being on the ground or
excited energy levels, correspondingly.

Inserting the expressions (\ref{1})-(\ref{3}) into the quantum mechanical
determination of atom momentum 
\begin{equation}
\left\langle p\right\rangle =\int \Psi ^{*}\widehat{p}\Psi d\overrightarrow{%
\rho }dz,\text{ \quad }\int \Psi ^{*}\Psi d\overrightarrow{\rho }dz=1
\label{4}
\end{equation}
and doing the standard transformations we arrive to 
\begin{equation}
\left\langle p\right\rangle =\int \left| a(p,t)\right| ^2pdp+\int \left|
b(p,t)\right| ^2pdp.  \label{5}
\end{equation}
The first member specifies the contribution of the ground energy level in
total momentum, 
\begin{equation}
\left\langle p\right\rangle _g=\int \left| a(p,t)\right| ^2pdp.  \label{5'}
\end{equation}
Accordingly, the second member specifies the excited level's contribution, 
\begin{equation}
\left\langle p\right\rangle _e=\int \left| a(p,t)\right| ^2pdp  \label{5"}
\end{equation}
Both these momenta are time-dependent and their change for $t$ time of
interaction will be 
\begin{mathletters}
\begin{eqnarray}
\left\langle \Delta p\right\rangle _g &=&\int \left( \left| a(p,t)\right|
^2-\left| a(p,0)\right| ^2\right) pdp,  \label{6a} \\
\left\langle \Delta p\right\rangle _e &=&\int \left( \left| b(p,t)\right|
^2-\left| b(p,0)\right| ^2\right) pdp.  \label{6b}
\end{eqnarray}
When the atom interacts with the travelling wave, the coefficient $a(p,t)$
of ground level relates with the coefficient $b(p+\hbar k,t)$ of excited
level (spontaneous transitions are not taken into account) (Fig. \ref{fig11}%
). As a result we get 
\end{mathletters}
\begin{equation}
\left| a(p,t)\right| ^2+\left| b(p+\hbar k,t)\right| ^2=const=\left|
a(p,0)\right| ^2+\left| b(p+\hbar k,0)\right| ^2.  \label{7}
\end{equation}
By this relation we can connect $\left\langle \Delta p\right\rangle _g$ with 
$\left\langle \Delta p\right\rangle _e$: 
\begin{eqnarray}
\left\langle \Delta p\right\rangle _e &=&\int \left( \left| b(p+\hbar
k,t)\right| ^2-\left| b(p+\hbar k,0)\right| ^2\right) \left( p+\hbar
k\right) d\left( p+\hbar k\right) =  \label{8} \\
&=&-\int \left( \left| a(p,t)\right| ^2-\left| a(p,0)\right| ^2\right)
\left( p+\hbar k\right) d\left( p+\hbar k\right) =  \nonumber \\
&=&-\left\langle \Delta p\right\rangle _g+\hbar k\int \left( \left|
a(p,t)\right| ^2-\left| a(p,0)\right| ^2\right) dp=-  \nonumber \\
&=&-\left\langle \Delta p\right\rangle _g+\hbar k\text{ }\Delta n_g, 
\nonumber
\end{eqnarray}
where $\Delta n_g=-\Delta n_e=$ $\int \left( \left| a(p,t)\right| ^2-\left|
a(p,0)\right| ^2\right) dp=-\int \left( \left| b(p,t)\right| ^2-\left|
a(p,0)\right| ^2\right) dp$ is the change of ground level's population, or
which is the same, the population change $\Delta n_e$ of excited level with
the opposite sign (see \ref{16'}). From the equality of the first and last
parts of (\ref{8}) directly follows the well known inequality between the
momentum of photon and total atom: 
\begin{equation}
\left\langle \Delta p\right\rangle =\left\langle \Delta p\right\rangle
_g+\left\langle \Delta p\right\rangle _e=\hbar k\text{ }\Delta n_g\leq \hbar
k\text{ ;}  \label{9}
\end{equation}
Let, nevertheless, note that this 
\mbox{$<$}%
\mbox{$<$}%
one photon demarcafation%
\mbox{$>$}%
\mbox{$>$}%
pertains to the total momentum of the atom, but not to the momentum of
ground and excited levels, separately. Their changes, in accordance with (%
\ref{6a}) and (\ref{6b}), in principle, may be arbitrary, depending on
distributions of $\left| a(p,t)\right| ^2-\left| a(p,0)\right| ^2$ and $%
\left| b(p,t)\right| ^2-\left| b(p,0)\right| ^2$ in momentum space . From
the expressions (\ref{6a}) and (\ref{6b}) also is obvious that for $%
\left\langle \Delta p\right\rangle _g$ $\left( \left\langle \Delta
p\right\rangle _e\right) $ to get great values, is necessary the
distribution of $\left| a(p,t)\right| ^2-\left| a(p,0)\right| ^2$ (of $%
\left| b(p,t)\right| ^2-\left| b(p,0)\right| ^2$) to be strictly
non-symmetric, relative to the replacement $p\rightarrow -p$ and to have a
gathering in the range of great values of $\left| p\right| $.

And now let's show that one photon absorption/ emission process in the field
of travelling wave really allows a behavior, mentioned above. The
Hamiltonian of the system, in dipole approximation, may be written as 
\begin{equation}
\widehat{H}=\widehat{H_0}-\widehat{d}E(t,z),  \label{10}
\end{equation}
where $\widehat{H_0}$ is the free atom Hamiltonian, and $\widehat{d}$ is the
dipole moment operator and 
\begin{equation}
\overrightarrow{E}(t,z)=\frac{\overrightarrow{E}}2\exp (ikz-i\omega t)+c.c,%
\text{ \quad }t>0  \label{11}
\end{equation}
is the electric field, whose $\omega $ frequency is equal to the $\omega _0$
frequency of Bohr transition.

From Schrodinger equation for $A(t,z)$ and $B(t,z)$ amplitudes we arrive to 
\begin{mathletters}
\begin{eqnarray}
i\frac{\partial A(t,z)}{\partial t} &=&-\nu \exp (-ikz)B(t,z),  \label{12a}
\\
i\frac{\partial B(t,z)}{\partial t} &=&-\nu \exp (ikz)A(t,z),  \label{12b}
\end{eqnarray}
the Rabi-solutions of which are \cite{Landau} 
\end{mathletters}
\begin{mathletters}
\begin{eqnarray}
A(z,t) &=&A(z,0)\cos \nu t+iB(z,0)\exp (-ikz)\sin \nu t,  \label{13a} \\
B(z,z) &=&B(z,0)\cos \nu t+iA(z,0)\exp (ikz)\sin \nu t,  \label{13b}
\end{eqnarray}
where $\nu =dE/2\hbar $ represents the Rabi frequency, $d=\left\langle
\varphi _a\mid \widehat{d}\mid \varphi _b\right\rangle $.

Performing $\chi (P)$-expansion (see (3)) in (\ref{13a}) and (\ref{13b}), we
obtain 
\end{mathletters}
\begin{mathletters}
\begin{eqnarray}
a(p,t) &=&a(p,0)\cos \nu t+ib(p+\hbar k,0)\sin \nu t,  \label{14a} \\
b(p,t) &=&b(p,0)\cos \nu t+ia(p-\hbar k,0)\sin \nu t,  \label{14b}
\end{eqnarray}
At first, it is readily verified that if the atom is on one of energy levels
before the interaction, the extraordinary things doesn't take place. Really,
if for example $b(p,0)=0$, then 
\end{mathletters}
\begin{mathletters}
\begin{eqnarray}
\left\langle \Delta p\right\rangle _g &=&(\cos ^2\nu t-1)\int \left|
a(p,0)\right| ^2pdp=(\cos ^2\nu t-1)\text{ }\left\langle P\right\rangle
_g\mid _{t=0}  \label{15a} \\
\left\langle \Delta p\right\rangle _e &=&(1-\cos ^2\nu t)\text{ }\left[
\left\langle P\right\rangle _g\mid _{t=0}+\hbar k\right] ,  \label{15b}
\end{eqnarray}
that is the contribution of momentum per energy level evolves periodically
and this is the evolution merely caused by periodic exchange of population
between the energy levels (posed by the term $(1-\cos ^2\nu t))$. Note also
that in conditions under consideration the momentum distributions coincide
with each other with a shift $\hbar k$: $b(p,t)=i$ $a(p,t)$ tg$\nu t$, as
was mentioned in Introduction.

The situation is totally diverse, if the atom is initially in superposition
state of ground and excited levels, because now the initial momentum
distributions on the ground and excited levels are not under necessity to be
identical with shift $\hbar k$: $b(p,0)\neq \alpha $ $a(p-\hbar k,0)$ in
general ($\alpha $ is some constant, independent of $p$). Then, it follows
unavoidably from (\ref{14a}) and (\ref{14b}) that the optical transition,
besides the changes on the energy levels'populations, leads also to periodic
evolutions in the form of momentum distributions there. Herewith the atomic
amplitudes $a(p,t)$ and $b(p,t)$ aren't mutually proportional (with any
constant shift).

To wash out the contributions, appropriate to evolution of the energy level
populations, let introduce a pair of new quantities, $\overline{p_g}$ and $%
\overline{p_e}$, which would be scaled in units of level-populations $n_g$
and $n_e$ respectively: 
\end{mathletters}
\begin{equation}
\overline{p_g}=\left\langle p\right\rangle _g/n_g,\text{ \quad }\overline{p_e%
}=\left\langle p\right\rangle _e/n_e  \label{16}
\end{equation}
\begin{equation}
n_g=\int \left| a(p,t)\right| ^2dp,\text{ \quad }n_g=\int \left|
b(p,t)\right| ^2dp  \label{16'}
\end{equation}
Since these new quantities are already independent on level populations,
their possible evolutions would be stipulated by form-deformations in the
energy levels momentum distributions. Afterwards we will call them mean
momenta of ground energy level ($\overline{p_g}$) and of excited energy
level ($\overline{p_e}$) correspondingly.. Thereby the total momentum of
atom, in addition to (\ref{5}), can be presented in the following more
convenient form: 
\begin{equation}
\left\langle p\right\rangle =n_g\overline{p_e}+n_e\overline{p_g}  \label{17}
\end{equation}
These mean momenta, $\overline{p_g}$ and $\overline{p_e}$, remain constant,
of course, if the atom is initially on one of the energy levels. They remain
constant also when the initial distributions are mutually proportional with
constant shift $\hbar k$: 
\begin{equation}
b(p,0)=\alpha \text{ }a(p-\hbar k,0)  \label{18}
\end{equation}
Really, putting (\ref{18}) in relations (\ref{14a}) and (\ref{14b}) making
the obvious substitutions , we arrive to 
\[
\overline{p_g}=\frac{\left| \cos \nu t-i\alpha \sin \nu t\right| ^2\int
\left| a(p,0)\right| ^2pdp}{\left| \cos \nu t-i\alpha \sin \nu t\right|
^2\int \left| a(p,0)\right| ^2dp}=\overline{p_g}\mid _{t=0} 
\]
for ground energy level, and $\overline{p_e}=\overline{p_e}\mid _{t=0}$ for
excited energy level. In these circumstances in (\ref{17}) the time
evolution exhibit only energy level populations $n_e$ and $n_g$.

In the general case, nevertheless, the evolution of state is due to
interference of non-similarly distributed amplitudes, the atomic amplitudes
distributions on energy levels aren't proportional with each other yet and,
as a consequence, mean momenta $\overline{p_e}$ and $\overline{p_g}$ get
temporal evolution, too.

For acquisition of more concrete and quantitative results, let's note that
for the intentions of atom optics and interferometry\cite{Adams} the
coherent scattering of atoms in the resonant field of standing wave is the
routine for preparation of large spreaded momentum distribution. The
probability amplitudes, prepared in a such way, can't satisfy the 
\mbox{$<$}%
\mbox{$<$}%
undesirable%
\mbox{$>$}%
\mbox{$>$}%
condition (\ref{18}) in principle, since in the field of standing wave . as
is well known, any state with momentum $p$ on one energy level is connected
with the two-states on other energy level with momenta $p-\hbar k$ and $%
p+\hbar k$ simultaneously. Therefore, any atom prepared by means of resonant
Kapitza-Dirac effect, during its later interaction with the travelling wave,
ought to implicitly change the momentum distributions on energy levels with
following to it above mentioned consequences.

\section{Preparation of Superpositional States on Atomic Ground and Excited
Levels by Means of Scattering in the Field of Resonant Standing Wave}

Let before the interaction with the travelling wave, during time $\tau _s$,
the atom had coherent interaction with the resonant ($\omega =\omega _0$)
standing wave \cite{Cook}. We restrict ourselves to the relatively simple
case, when the interaction proceeds by the well known scheme of mutually
orthogonal atom-standing wave beams. Moreover, the Raman-Nath approximation
will be employed, which permit to put out from the problem at hand the
kinetic energy term in the Hamiltonian (note that the kinetic energy term
has not been included into (\ref{10}) too). Although the scheme of
calculation is well known and presented in details (see, for example, in 
\cite{Adams}, \cite{Cook}, we find it convenient to give an account of main
intermediate formulas, too.

To describe the interaction in the preparing standing wave, the electric
field (\ref{11}) in the Hamiltonian (\ref{10}) must be exchanged by 
\begin{equation}
E(t,z)=E_s\cos kz\exp (-i\omega t)+c.c,\text{ }-\tau _s\leq t\leq 0.
\label{19}
\end{equation}

In consequence, the atomic amplitudes $A_s(z,t)$ and $B_s(z,t)$ have to
fulfill (\ref{12a},\ref{12b})-type equations where the following
replacements must be performed: $\nu \rightarrow 2\nu _s=2dE_s/\hbar $
(which is mean Rabi frequency in the standing wave), $exp(\pm
ikz)\rightarrow \cos kz$. Allowing that the atom has been on the ground
level before the interaction $(t<-\tau _s)$, we arrive to 
\begin{mathletters}
\begin{eqnarray}
A_s(z,t) &=&\cos (2\nu _s(t+\tau _s)\cos kz),\text{ }  \label{20a} \\
B_s(z,t) &=&i\sin (2\nu _s(t+\tau _s)\cos kz).\text{ }  \label{20b}
\end{eqnarray}
These amplitudes at the moment $t=0$, when the standing wave is turned off,
just present the initial amplitudes $A(z,0)$ and $B(z,0)$ for interaction
with the travelling wave (see the formulas (\ref{13a},\ref{13b}). Using
their $\chi (p)$-expantions \cite{Prudnikov}. 
\end{mathletters}
\begin{mathletters}
\begin{eqnarray}
A_s(z,0) &=&\cos (2\nu _s\tau _s\cos kz)=\sum_{m\text{ }=\text{ }-\infty
}^\infty i^{2m}J_{2m}(2\nu _s\tau _s)\exp (i2mkz),  \label{21a} \\
B_s(z,0) &=&i\sin (2\nu _s\tau _s\cos kz)=\sum_{m\text{ }=\text{ }-\infty
}^\infty i^{2m+1}J_{2m+1}(2\nu _s\tau _s)\exp (i(2m+1)kz),  \label{21b}
\end{eqnarray}
where $J_n(x)$ is Bessel function, for atomic center-of-mass motion
probability amplitudes $a(p,t)$ and $b(p,t)$ we get the following
expressions: 
\end{mathletters}
\begin{mathletters}
\begin{eqnarray}
a(2m\hbar k,t) &=&i^{2m}\left[ \cos \nu t\text{ }J_{2m}(2\nu _s\tau _s)-\sin
\nu t\text{ }J_{2m+1}(2\nu _s\tau _s)\right] ,  \label{22a} \\
b((2m+1)\hbar k,t) &=&i^{2m+1}\left[ \cos \nu t\text{ }J_{2m+1}(2\nu _s\tau
_s)+\sin \nu t\text{ }J_{2m1}(2\nu _s\tau _s)\right] ,  \label{22b} \\
a((2m+1)\hbar k,t) &=&b(2m\hbar k,t)=0  \label{22c}
\end{eqnarray}
The superposition state, created as a result of interaction with the
standing wave, present the discrete mainfields of states, where the space
between the adjacent values of momentum is $2\hbar k$, herewith the
mainfields for ground and excited levels are totally shifted with respect to
each other by $\hbar k$ (the half of $2\hbar k$) \cite{Cook}.

The formulas (\ref{22a})-(\ref{22c}) contain explicitly the seeking result
about the evolution of momentum distributions. To exhibit this evolution,
let first note that the initial momentum distribution for both energy levels
are symmetric relative to value $p=0$. Really, they are specified by $%
i^{2m}J_{2m}(.)$ and $i^{2m+1}J_{2m+1}(.)$ functions for ground and excited
energy levels respectively and are symmetric, relative to $2m\rightarrow -2m$%
, $2m+1\rightarrow -(2m+1)$ transformations, that is just relative to value $%
m=0$ $(p=0)$. This symmetry signifies that the momentum of each energy level
(as the incremental (\ref{5}), as the mean (\ref{17}) values) is zero \cite
{Cook} before the interaction with the travelling wave. Nevertheless the
symmetry breaks under the 
\mbox{$<$}%
\mbox{$<$}%
influence%
\mbox{$>$}%
\mbox{$>$}%
of travelling wave: one photon absorption/emission process, in accordance
with (\ref{22a}) and (\ref{22b}), gives the beginning of asymmetric
transformations in the form of momentum distributions, periodically running
in opposite directions for ground and excited energy levels.

A typical form of initial distributions and the following redistributions
(due to single-photon process) are depicted on Fig. \ref{fig22} and Fig. \ref
{fig33} for ground and excited energy levels consequently. Single-photon
large-scale changes are apparent.

Now let notice that in conditions of Fig. \ref{fig22} and Fig. \ref{fig33}
we get almost one-side distributions: the translational states with $n>0$
for ground energy level only, and the translational states with $n<0$ for
excited level only. So, the state of total atom has been explited into two
sub-groups, where one sub-group presents ground-level atoms with negative
values of momentum, and the second sub-group presents vice-versa, the
excited-level atoms with negative values of momentum. Of course, this is a
Stern-Gerlach type splitting. That is, one-photon optical transition
implements the resonant Kapitza-Dirac splitting into the Stern-Gerlach type
splitting.

The phenomenon of one-photon coherent accumulation of momentum on energy
levels (OP-CAMEL), may get some expansion in exhibition, if the initial
momentum distributions would be taken in asymmetric form. This kind of
distributions also can be obtained by the standing wave, but if some
travelling wave is previous to it \cite{Romanenko}. Such sequence of pulses
is got , if the standing wave is formed by means of reflection of a laser
pulse from the distanted (from atomic beam) mirror (see, for example \cite
{Grinchuk},). To avoid the overloading of the text we aren't going to bring
formulas and the behavior of the OP-CAMEL in these conditions will be given
only by some graphs. In Fig. \ref{fig44} and Fig. \ref{fig55} is presented
the case when the momentum distribution in resonant Kapitza-Dirac splitting
is maximum asymmetric. As is seen from the graphs, in this special case
OP-CAMEL appears already as a accumulation of asymmetry on one (ground)
energy level for account of its suppressing on the other (excited) energy
level.

\section{Time Evolution of Mean Momentum On Ground and Excited Energy Levels
in the Field of Travelling Wave}

Let us now return to the preparation only by the standing wave and discuss
the evolution of momenta $\overline{p_g}$ and $\overline{p_e}$. By means of
expressions for the quantities defining $\overline{p_g}$ and $\overline{p_e}$
(see (\ref{16}), (\ref{16'}) and (\ref{5'}), (\ref{5"})) we will have 
\end{mathletters}
\begin{eqnarray}
\left\langle p\right\rangle _g &=&\hbar k\sum_{m\text{ }=\text{ }-\infty
}^\infty 2m\left[ \cos \nu t\text{ }J_{2m}(u)-\sin \nu t\text{ }%
J_{2m+1}(u)\right] ^2=  \label{23} \\
&=&-\hbar k\left[ \frac{1-J_0(2u)}2\sin ^2\nu t+\frac{u-J_1(2u)}4\sin 2\nu
t\right]  \nonumber
\end{eqnarray}
\begin{eqnarray}
n_g &=&\sum_{m\text{ }=\text{ }-\infty }^\infty \left[ \cos \nu t\text{ }%
J_{2m}(u)-\sin \nu t\text{ }J_{2m+1}(u)\right] ^2=  \label{23'} \\
&=&\frac 12+\frac{J_0(2u)}2\cos 2\nu t-\frac{J_1(2u)}2\sin 2\nu t\text{ } 
\nonumber
\end{eqnarray}
on the ground energy level, and 
\begin{eqnarray}
\left\langle p\right\rangle _e &=&\hbar k\sum_{m\text{ }=\text{ }-\infty
}^\infty (2m+1)\left[ \cos \nu t\text{ }J_{2m+1}(u)+\sin \nu t\text{ }%
J_{2m+1}(u)\right] ^2=  \label{24} \\
&=&\hbar k\left[ \frac{1+J_0(2u)}2\sin ^2\nu t+\frac{u+J_1(2u)}4\sin 2\nu
t\right] ,  \nonumber
\end{eqnarray}
\begin{eqnarray}
n_e &=&\sum_{m\text{ }=\text{ }-\infty }^\infty \left[ \cos \nu t\text{ }%
J_{2m+1}(u)+\sin \nu t\text{ }J_{2m}(u)\right] ^2=  \label{24'} \\
&=&\frac 12-\frac{J_0(2u)}2\cos 2\nu t+\frac{J_1(2u)}2\sin 2\nu t=1-n_g\text{
}  \nonumber
\end{eqnarray}
on the excited energy level. Here $u=2\nu _s\tau _s.$ The last forms of (\ref
{23})-(\ref{24'}) are got by using for formulas of summations of Bessel
functions \cite{Ref4}). As $\left\langle p_g\right\rangle \mid _{t=0}=0$, $%
\left\langle p_e\right\rangle \mid _{t=0}=0$, (the same for $\overline{p}_g$
and $\overline{p}_e$), then their values at any next moment t present
simultaneously their changes for the time $t$: $\left\langle \Delta
p\right\rangle _g=\left\langle p\right\rangle _g$, $\left\langle \Delta
p\right\rangle _e=\left\langle p\right\rangle _e$. In Fig. \ref{fig66} and 
\ref{fig77} are given the temporal evolutions of momenta, corresponding to
each energy level, while they are interacting with travelling (accumulating)
wave. Population changes, which also are responsible for the evolutions in
general, are depicted in figures by dashed lines. In presented case the
populations are constant in practice, which follows from (\ref{23'}) and (%
\ref{24'}) too, if we take into account $J_{0,1}(2u)<<1$ for $u>>1$. And,
finally, the temporal evolution of mean momenta $\overline{p}_g$ and $%
\overline{p}_e$ conditioned solaly by redistributions of momentumon the
energy levels, is shown already in Fig. \ref{fig88} and \ref{fig99}. The
parameters of the preparing standing wave are the same as in Fig. \ref{fig22}%
, where the distance between left-hand and right-hand maximums (the width of
momentum distribution) is about $70$ $\hbar k$. Such magnitudes for the
resonant Kapitza-Dirac splitting are totally in limits of experimental
realizations \cite{Ref3}.

Note also, that the comparison of the deviation of $\overline{p}_g$ or $%
\overline{p}_e$ (from the Fig. \ref{fig88} and \ref{fig99}) and the width of
momentum distribution (from the \ref{fig22}) shows the same order of
magnitude for them. Since the width of momentum distribution has multiphoton
nature (created by means of multiphoton process of reemission of photons
from one wave into the counterpropagating one), the large -scale variations
in OP-CAMEL may be called as 
\mbox{$<$}%
\mbox{$<$}%
multiphoton%
\mbox{$>$}%
\mbox{$>$}%
.

Multiphoton OP-CAMEL\ manifests itself in (\ref{23}), (\ref{24}) formulas in
following way. When the initial momentum distribution is sufficiently
widespread, that is $\Delta p\gg \hbar k,$ then $2\nu _s\tau _s\gg 1$ (from
the theory of resonant Kapitza-Dirac effect the connection between the
momentum width $\Delta P$ and the number of Rabi-flops $2\nu _s\tau _s$ is $%
\Delta p\approx 2\nu _s\tau _s\hbar k$ )). Under this condition the members $%
-\frac 14$ $\hbar k$ $u\sin \nu t$ in (\ref{23}) and $-\frac 14$ $\hbar k$ $%
u\sin \nu t$ in (\ref{24}), being proportional to $2\nu _s\tau _s>>1$, stand
out as the previal terms and just present the multiphoton OP-CAMEL.

\section{Summary}

A simple theoretical consideration of optical transition in general
conditions, when the atom initially has in superposition state of ground and
excited energy levels different momentum distributions, shows that one
one-photon optical transition leads to radical asymmetric changes in
momentum distributions on each energy level. Saying in images, in general
one photon changes the mean momentum of energy level more than photon's own
momentum.

In important case, when the preliminary superposition state of the atom is
being prepared by coherent scattering at the resonant standing wave, the
phenomenon can be presented as a transition from resonant Kapitza-Dirac
splitting of atomic translational states to Stern-Gerlach type splitting.
This is schematically depicted in Fig. \ref{fig1010}.

Finally, let's give an account of some remarks about the possibility of
experimental observation of phenomenon. Firs let's notice that the 
\mbox{$<$}%
\mbox{$<$}%
non-optical%
\mbox{$>$}%
\mbox{$>$}%
methods, which registrate the total atom (for example 
\mbox{$<$}%
\mbox{$<$}%
hot-wire%
\mbox{$>$}%
\mbox{$>$}%
method), can't be used for this purpose, because the phenomenon deals with
each individual energy level; the momentum distribution of total atom
doesn't change or which will be more right, it changes only in one-photon
momentum limits.

It is preferable to use registration methods, that will deal only with one
of resonantly connected energy levels, such as the adjacent optical
transitions. Then the phenomenon will appear itself as a pronounced
asymmetry in profile of Doppler broadening, relative to Bohr frequency.
Another possibility we see in using of long-living energy levels, so long
that the atomic translational states can be distinguished in space till the
spontaneous emission (C zone in Fig. \ref{fig1010}). In this case the
space-sensitive schemes of spontaneous emission collection or probe pulse
absorption will bring to desirable result.

This research is supported by the Armenian State Scientific Foundation under
Grant No 96-901.

\begin{figure}[tbp]
\caption{The scheme of interaction of a two-level atom with the resonant
travelling wave. The initial state of the atom is prepared by the standing
wave and presents a manifold of discrete translational states per each
energy level.}
\label{fig11}
\end{figure}
\begin{figure}[tbp]
\caption{Probability distribution $W_{ground}^{(n)}$ of definite-momentum
states on ground level prepared by the interaction with the standing wave.
Integer $n$ determines the value of momentum $($ $p=n\hbar k)$. Momentum
distribution before the travelling wave was symmetric. The chosen parameters
are $2\nu _s\tau _s=40,$ consequently $\left| \frac{A(-\tau _s)}{\chi (P_0)}%
\right| ^2=1,\left| \frac{B(-\tau _s)}{\chi (P_0)}\right| ^2=0,\nu t=20.$}
\label{fig22}
\end{figure}
\begin{figure}[tbp]
\caption{Probability distribution $W_{excited}^{(n)}$ of definite-momentum
states in excited level after the interaction with the standing and
travelling waves. Integer $n$ determines the value of momentum $(P=$ $%
p+n\hbar k)$. Momentum distribution before the travelling wave was
symmetric. The chosen parameters are $\nu _s\tau _s=20,$ consequently $%
\left| \frac{A(-\tau _s)}{\chi (P_0)}\right| ^2=1,\left| \frac{B(-\tau _s)}{%
\chi (P_0)}\right| ^2=0.$ }
\label{fig33}
\end{figure}
\begin{figure}[tbp]
\caption{Probability distribution $W_{ground}^{(n)}$ of definite-momentum
states on ground level prepared asymmetrically by theby the interaction with
the standing wave. Integer $n$ (as in Fig. 2 and 3) determines the value of
momentum $($ $p=n\hbar k)$. Momentum distribution before the travelling wave
was symmetric. The chosen parameters are $2\nu _s\tau _s=40,$ consequently $%
\left| \frac{A(-\tau _s)}{\chi (P_0)}\right| ^2=1/2,\left| \frac{B(-\tau _s)%
}{\chi (P_0)}\right| ^2=1/2,\nu t=20.$}
\label{fig44}
\end{figure}
\begin{figure}[tbp]
\caption{Probability distribution $W_{excited}^{(n)}$ of definite-momentum
states in excited level prepared asymmetrically by the interaction with the
standing and travelling waves. The chosen parameters are the same as in Fig.
4$.$}
\label{fig55}
\end{figure}
\begin{figure}[tbp]
\caption{Temporal behavior of momentum of translational motion for ground
energy level. Time interval includes all parts of interaction: with
preparing-standing and final-travelling waves. All parameters have the same
values as in fig. 2.}
\label{fig66}
\end{figure}
\begin{figure}[tbp]
\caption{Temporal behavior of momentum of translational motion for excited
energy level.All parameters have the same values as in fig. 2.}
\label{fig77}
\end{figure}
\begin{figure}[tbp]
\caption{Temporal behavior of mean momentum of translational motion for
ground energy level. Time interval includes al parts of interaction: with
preparing-standing and final-travelling waves. All parameters have the same
values as in fig. 2.}
\label{fig88}
\end{figure}
\begin{figure}[tbp]
\caption{Temporal behavior of mean momentum of translational motion for
excited energy level. Time interval includes al parts of interaction: with
preparing-standing and final-travelling waves. All parameters have the same
values as in fig. 2.}
\label{fig99}
\end{figure}
\begin{figure}[tbp]
\caption{The ground level, definite momentum state of an atom (zone 1)
preliminary transforms into a superposition one by coherent interaction with
a resonant standing wave (zone 2). The next interaction with the travelling
wave leads to large-scale changes in atomic momentum distributions.}
\label{fig1010}
\end{figure}

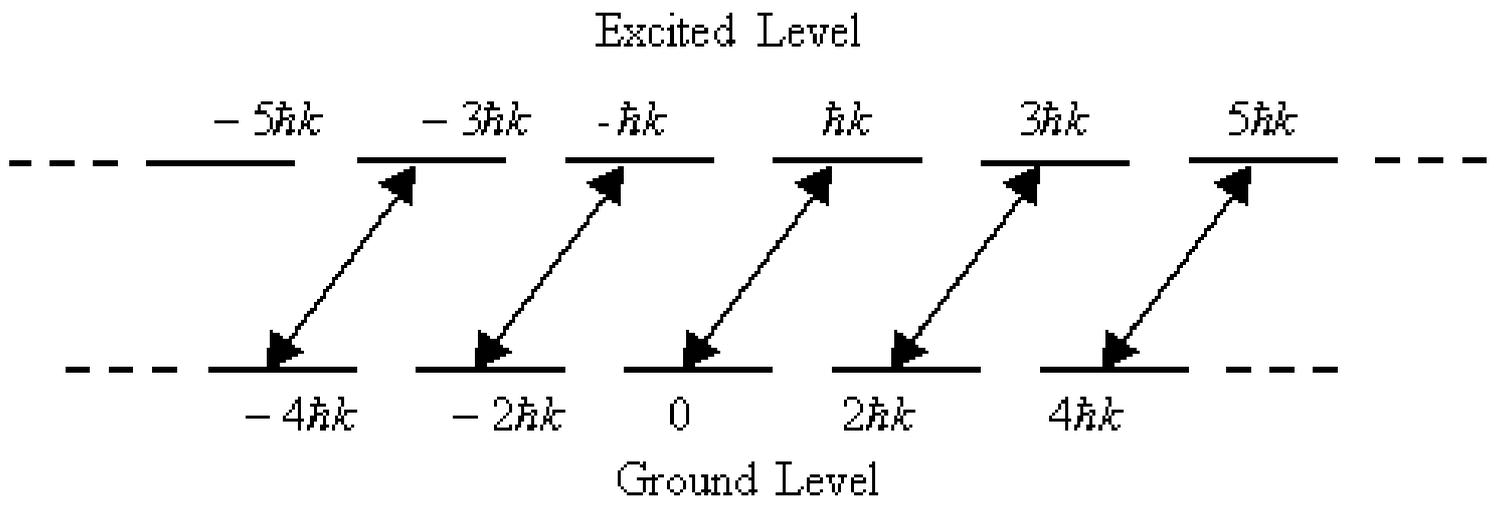
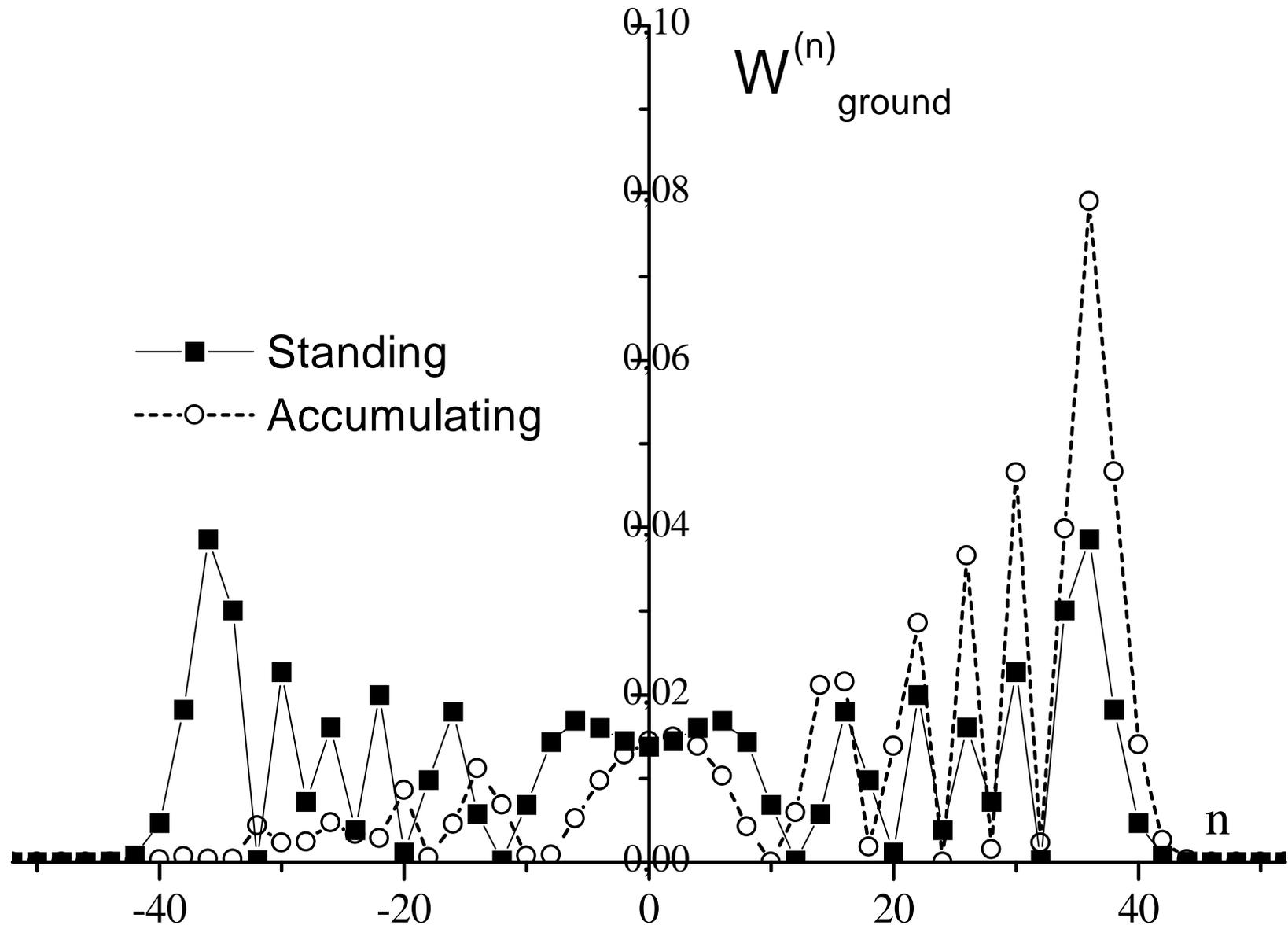
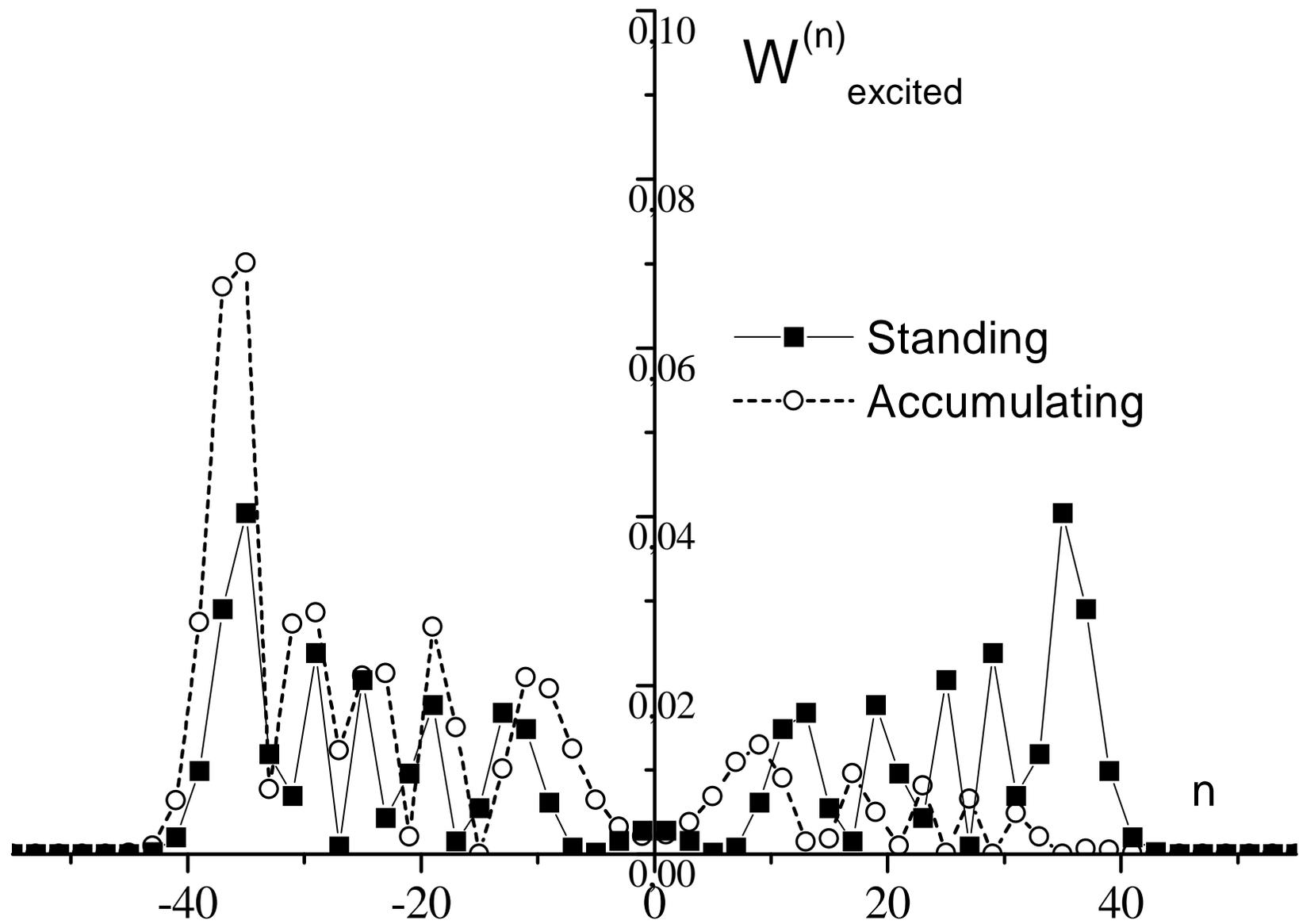
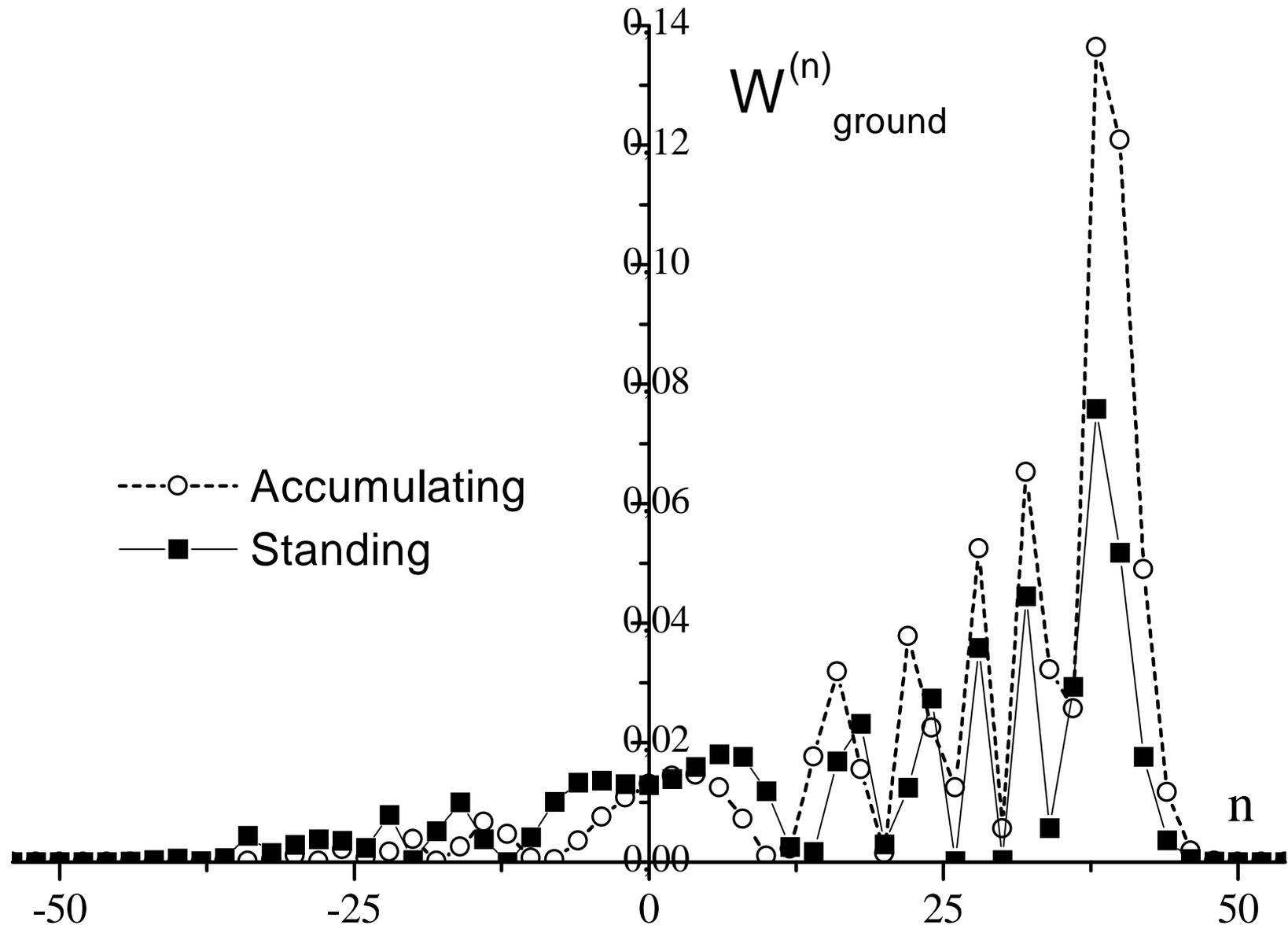
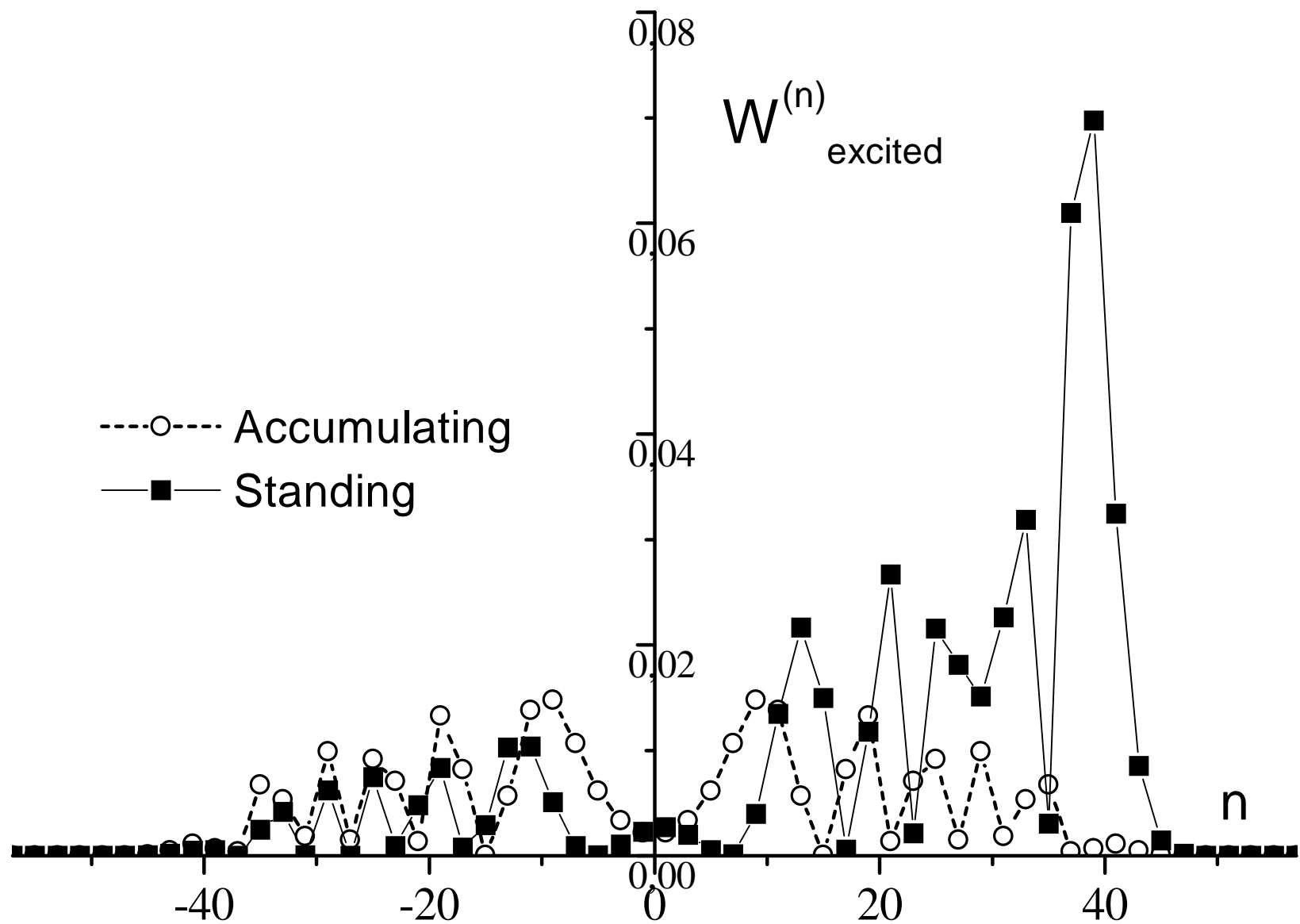
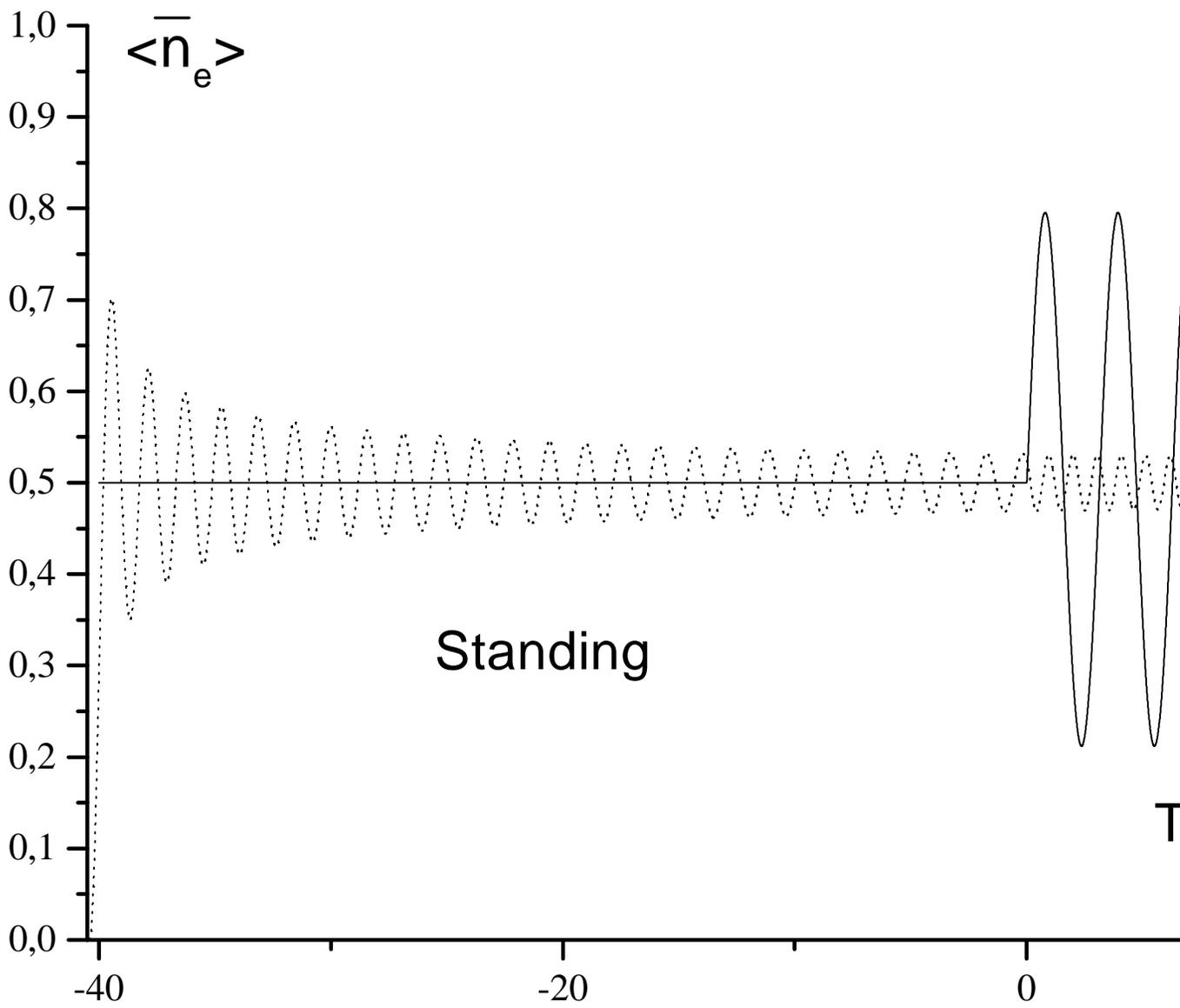
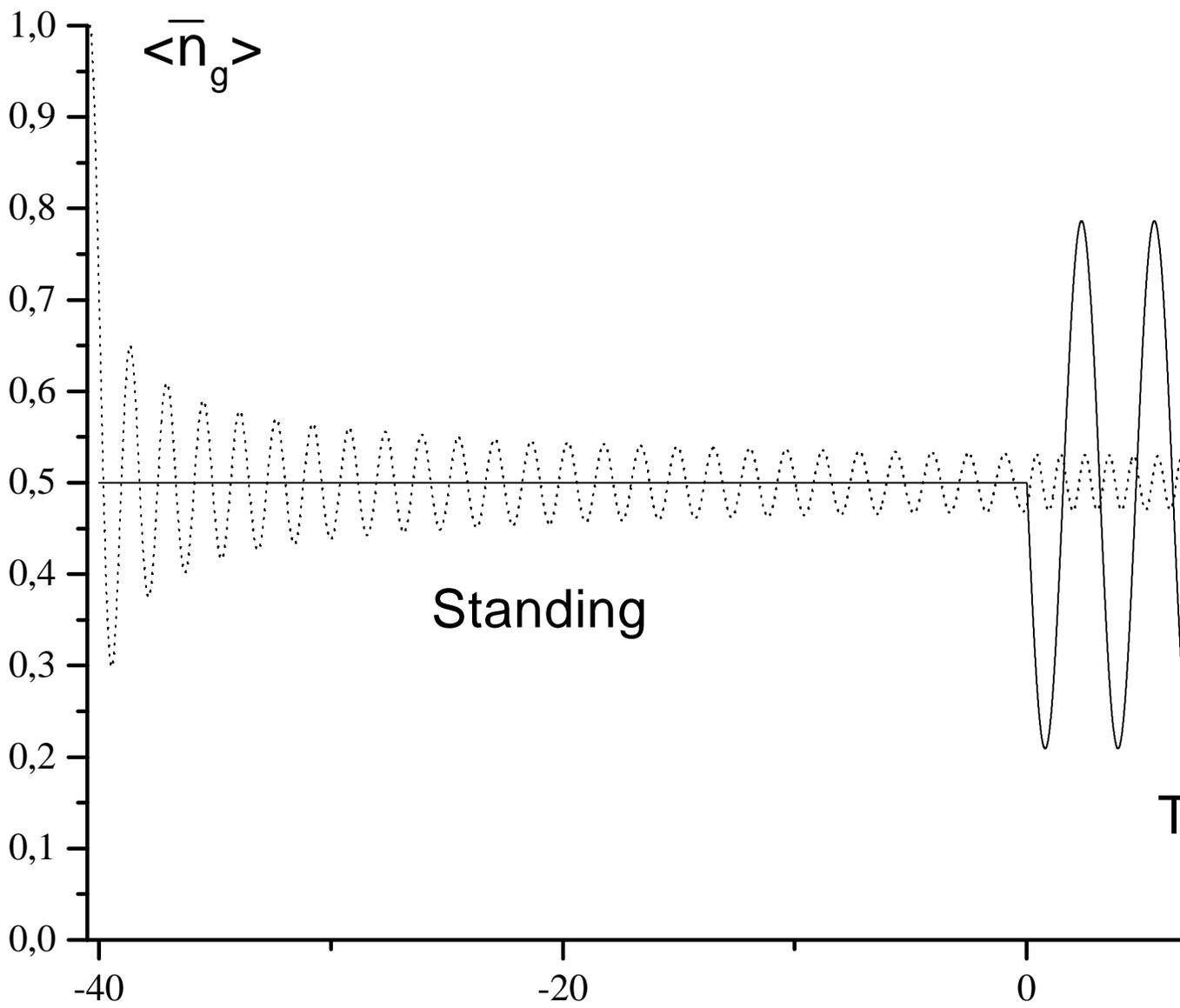
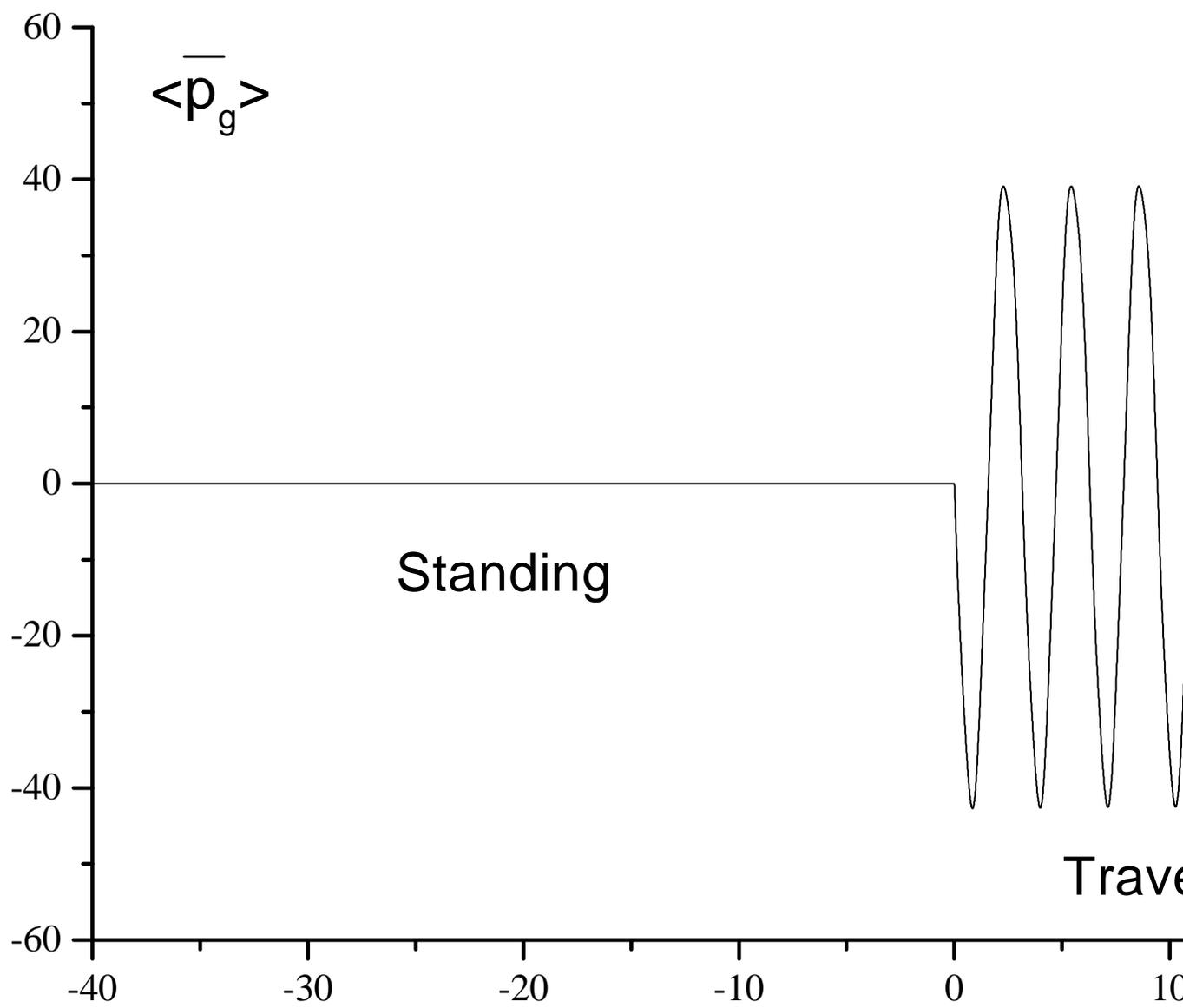
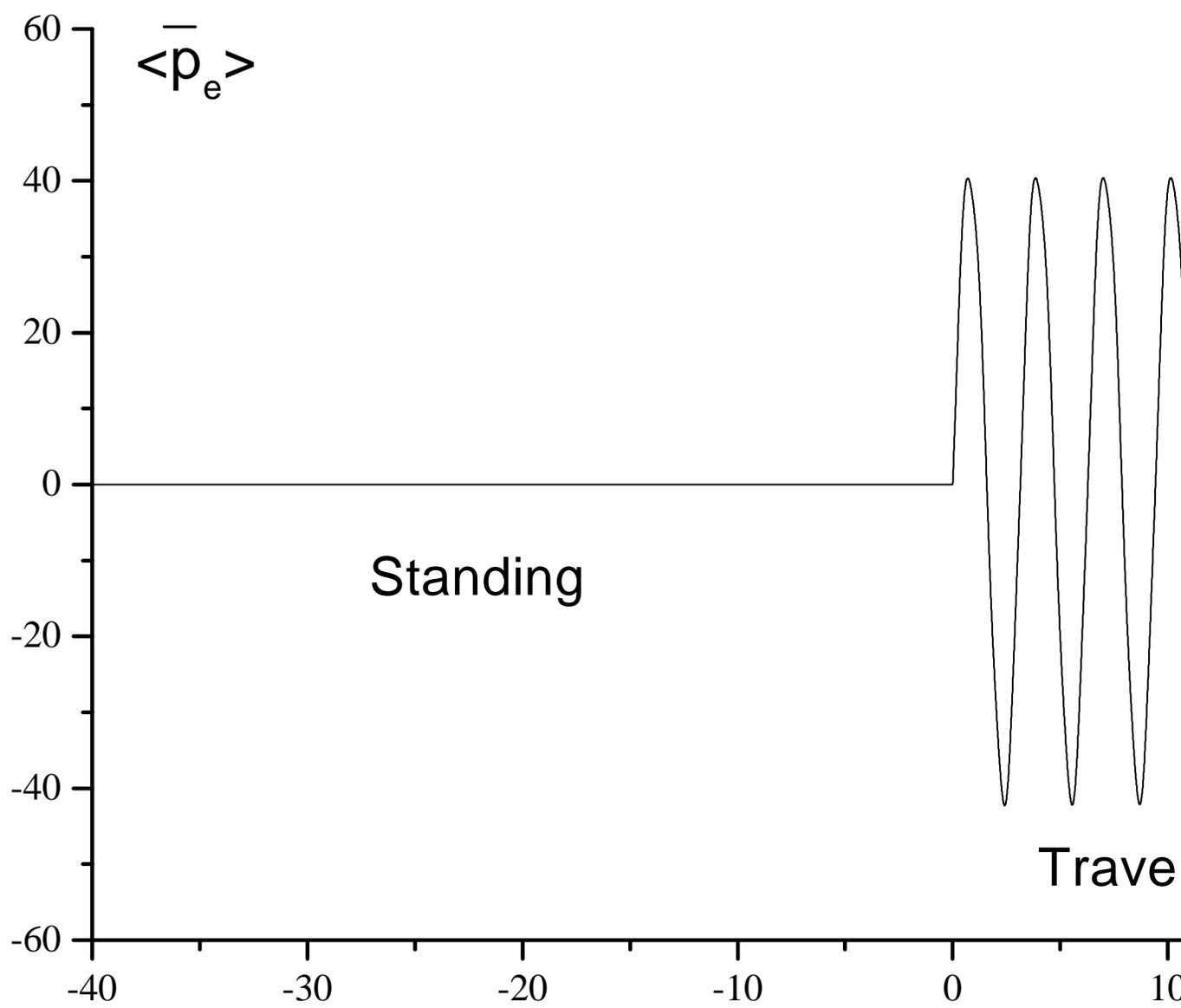
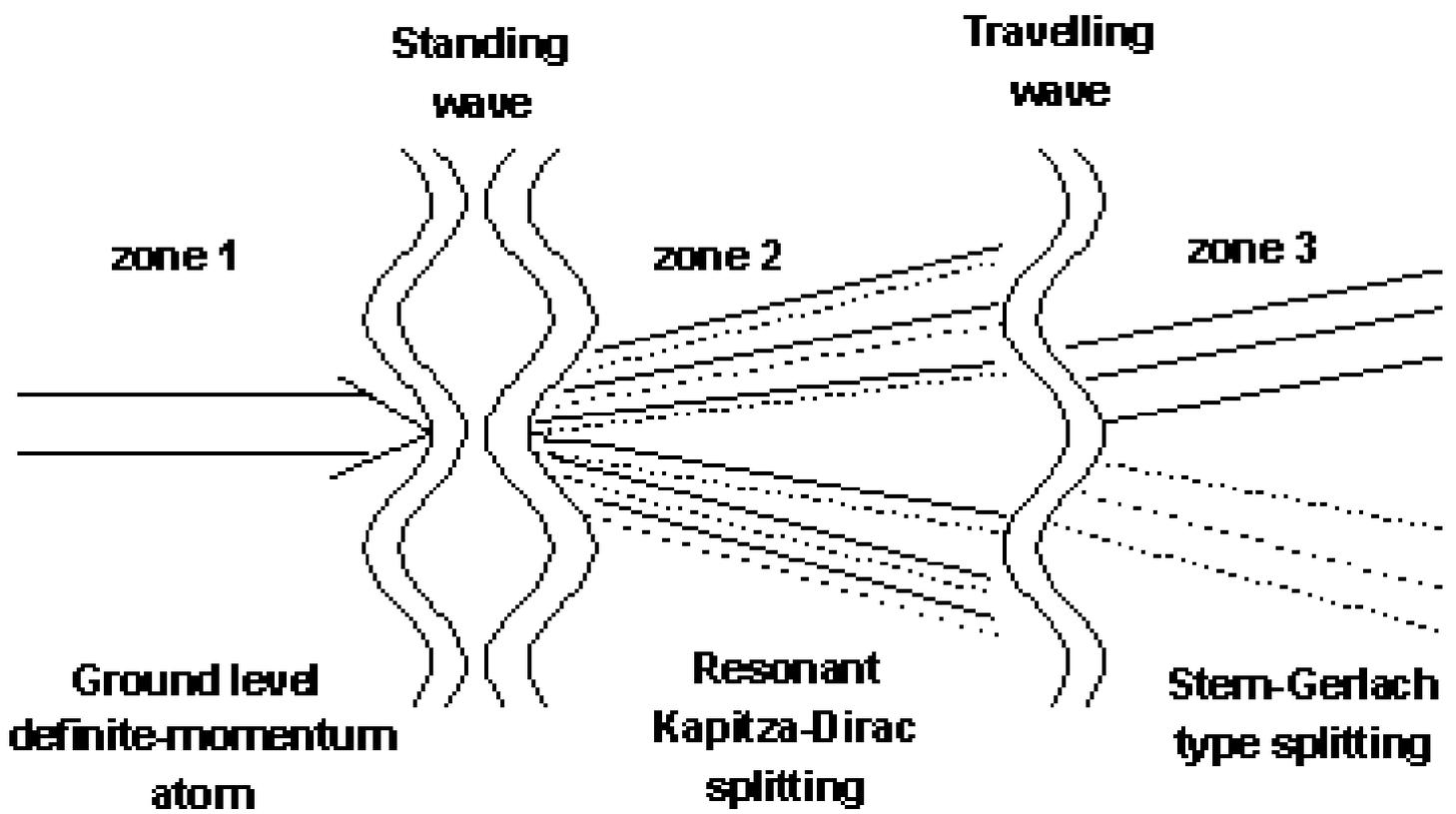
\end{document}